\documentclass{svproc}
\usepackage{url}
\usepackage{color,soul} 
\usepackage{tabularx}
\usepackage[table,xcdraw]{xcolor}

\usepackage{graphicx}
\usepackage{verbatim} 
\usepackage{wrapfig}
\usepackage{graphicx}
\usepackage{}

\begin{document}

\mainmatter              
\title{Statistical Analysis of ReRAM-PUF based Keyless Encryption Protocol Against Frequency Analysis Attack}

\titlerunning{Statistical Analysis of Keyless Encryption Protocol}  
%
\author{Dina Ghanai Miandaob \and Sareh Assiri\and
Joseph Mihaljevic\and Bertrand Cambou}
\authorrunning{Dina Ghanai et al.} 

\tocauthor{Dina Ghanai, Sareh Assiri,Joseph Mihaljevic, Bertrand Cambou}

\institute{Northern Arizona University, Flagstaff AZ 86011, USA,\\
\email{dinaghanai@nau.edu, sa2363@nau.edu, Joseph.Mihaljevic@nau.edu, Bertrand.Cambou@nau.edu}}

\maketitle              

\begin{abstract}
There has been a growing interest in fully integrating Physical Unclonable Function (PUF) for cryptographic primitives, or keyless encryption. Keyless primitives do not store key information during the entire encryption and decryption phase, providing full security against volatile and non-volatile memory attacks. The concept of keyless encryption using ReRAM PUF is a relatively new concept, and the security aspect of the protocol has not been tested yet. In this paper, we use statistical models to analyze the randomness of the protocol and its resistance against frequency attacks.
\end{abstract}

\section{Introduction}

Classical cryptography uses keys for encryption and decryption. Key generation, key exchange, and key storage are complex problems. Hackers   try to exploit the key to attack a system. Attacks based on differential power analysis can extract cryptographic keys during the encryption and decryption process. Quantum adversaries are targeting key extractions as well \cite{cambou2021post}. Internet of things (IoT) devices are being used more and more in everyday life. It has become an inevitable part of human life, which can also threaten one’s privacy since it might contain information in the communication that a third party should not be aware of. Hence, securing IoT devices is critical \cite{keshavarz2018towards}.

Another drawback of using keys for cryptography, however, is the large memory space allocated for key storage in devices. Most of the IoT devices  do not possess ample memory space due to cost constraints and limited power supply \cite{cambou2020cryptography,roman2011securing,baracaldo2016securing}. This means that the problem of longer secret keys and strong cryptography systems can be hard to implement in IoT devices.  These obstacles were a motivation to develop a keyless cryptography scheme.

Kumar \cite{kumar2003keyless} and Chandrasekaran \cite{chandrasekaran2015keyless} proposed a keyless authentication protocol, which uses a challenge-response pair to define the success or failure of authentication. A keyless  encryption scheme, which uses memristors technology, is developed by Bertrand Cambou et. al \cite{cambou2020cryptography}. This protocol uses a memristor PUF (see table \ref{tab:Abbreviations}) to create ciphertext without generating a key. In this paper, we explore the security of this approach. The protocol that has been developed uses the ReRAM PUF for encryption and decryption purposes. Also, the protocol uses a Random number generator and hash function to increase the level of randomness. With the help of RNG and hash function output, the protocol can visit particular cells in ReRAM PUF. The security of this approach will be compromised if, for any same plain text, the same cells of the ReRAM PUF are visited. Since there hasn’t been any study on the security of this approach, we studied the randomness of visiting cells using the same plaintext. Therefore, this study's question is how randomly every cell of ReRAM PUF is used for a given plaintext.

To analyze the randomness of the protocol of keyless encryption, we  used a binnomial error distrbution to estimate the probability of visiting each cell in the ReRAM PUF. 

This article is structured into five sections. Section 2 presents a brief background information about the keyless protocol that we used and descriptions of the tools such as PUF, memristors, and hash functions for the Random number generator (RNG), and hash functions.

Section 3 presents the main objectives of the work and the security threats. We worry about frequency analysis attacks; when encrypting the same plaintext multiple times, if we visit the same cells repeatedly, the hacker will use the frequency analysis technique to detect the plaintext, and the security will be compromised. So, this study tries to show if hackers can exploit frequency analysis against the keyless encryption scheme using the ReRAM PUF protocol. In addition, section 3 presents our methods of collecting data and how the data was labeled in the experiment.

Section 4 presents a statistical model that was used to test the probability of visiting each cell in the ReRAM PUF. We want to know how many times each cell was visited to indicate success or failure. Our goal was to determine whether specific cells were more or less likely to be visited on average, which could be exploited by hackers.

Section 5 presents the results of the statistical analysis. We provide tables, graphs, and results from a comparison among models.

\begin{table}[]
\centering
\caption{Abbreviations}
\label{tab:Abbreviations}
\resizebox{\textwidth}{!}{%
\begin{tabular}{|
>{\columncolor[HTML]{FFFFFF}}l |
>{\columncolor[HTML]{FFFFFF}}l |}
\hline
\textbf{PUF}    & Physical Unclonable Functions       \\ \hline
\textbf{SHA}    & Secure Hash Algorithms              \\ \hline
\textbf{MD}   & \begin{tabular}[c]{@{}l@{}}Message-Digest \\ (output of the hash function)\end{tabular} \\ \hline
\textbf{ReRAM}  & Resistive Random-Access Memory      \\ \hline
\textbf{IoT}    & Internet of things                  \\ \hline
\textbf{SMD}    & Short message-digest                \\ \hline
\textbf{LMD}    & Longer   message-digest             \\ \hline
\textbf{RNG}    & Random Number Generator             \\ \hline
\textbf{TRN}    & True random number                  \\ \hline
\textbf{PRN}    & Pseudo-random number                \\ \hline
\textbf{NIST} & National Institute of Standards and Technology                                          \\ \hline
\textbf{LOO-IC} & Leave one out information criterion \\ \hline
\end{tabular}%
}
\end{table}

\begin{figure}[ht]
\centering
\includegraphics[width=\linewidth]{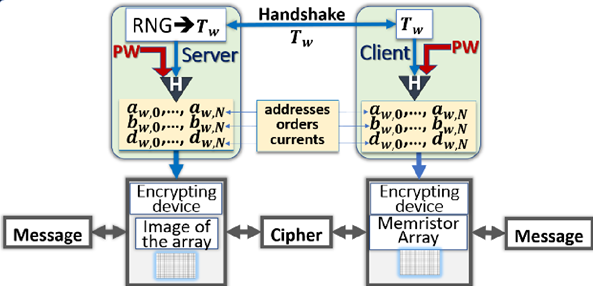}
\caption{The architecture schemes for keyless encryption with memristor PUF (ReRAM). The handshake means that a True Random Number share among sender and receiver before starting encryption and decryption. The ciphertext will be the outcome of reading the ReRAM PUF \cite{cambou2020cryptography}.}
\label{fig:Thearchitectureschemesforkeylessencryption}
\end{figure}

\section{Background}
A memristor is a combination of a memory and a transistor to generate a new type  of memory \cite{chua1971memristor,johnsen2012introduction,stanley2013we}. Resistive Random-Access Memory (ReRAM) works by changing the resistance across a die \cite{adam20173d}. It is often referred to as a memristor. In 2017, \cite{cambou2020puf} mentioned that “the injection of low currents  in cells of memristor arrays, can result in dissolvable conductive paths of vari able resistances, and can be exploited to design Physical Unclonable Functions (PUFs).”

Articles \cite{cambou2020cryptography} and \cite{zhu2020extended} describe how a memristor PUF is used to encrypt plaintext without using an encryption key. Through exploiting the features of memristor, the output of memristors becomes the cipher. After subjecting the current to a specific cell, the output will be the value of resistance. As shown in \cite{cambou2020cryptography} and
\cite{zhu2020extended} , resistance values with the current value can make ciphertext. Repeatedly visiting the same cell in the memristor can help hackers exploit frequency analysis attacks to extract the plaintext. According to \cite{cambou2020cryptography}, the ReRAM cells have been used  to generate a ciphertext. This paper is going to focus on studying the randomness  of visiting each cell in the memristor.

A keyless protocol based on ReRAM PUF uses three main tools: Random Number Generator (RNG), Hash function, and ReRAM PUF as shown in  \ref{fig:Thearchitectureschemesforkeylessencryption}. These tools depend on each other to build the keyless protocol. So, this section will give a brief description of the tools before explaining the keyless protocol steps \cite{cambou2021post}.

\subsection{Tools}
To run the protocol, several tools, including Random Number Generator (RNG), Hash function, and memristor PUF, are being used. In this section, we describe each tool and how it was used.

\textbf{Random Number Generator (RNG)}
The random number is the number that is generated randomly from one of the random number generators. The aim of the RNG is to get a unique number without repetition. There are two types of RNG, which are the true random number (TRN) and pseudo-random number (PRN). The TRN can be generated from the physical processes whereas    the PRN can be generated mathematically. In this study, we selected TRN because it is recommended by the National Institute of Standards and Technology (NIST). We did not need to test the RNG’s randomness because it already passes the NIST’s tests \cite{paar2009understanding}.

\textbf{Hash Function}
A hash function is defined as a one-way encryption which means that reversing the output of the hash will not help to retrieve the input. The hash function accepts arbitrary input and gives fixed output. The hash function has several common algorithms such as MD5, SHA-1, SHA-2, NTLM,    and LANMAN \cite{paar2009understanding} \cite{LSORainb88:online}. According to \cite{cambou2020cryptography} and \cite{zhu2020extended}, the SHA-3-512 was selected to  be used in the protocol implementation, which is one of the hash functions recommended by the NIST.

\textbf{Memristor PUF}
Physically Unclonable Functions (PUFs) are equivalent to human biometrics for a physical device. PUFs are hard to clone, hard to predict, and difficult to replicate; but, they  have a repeatable behavior. PUF targets the nanoscale device parametric variations to create unclonable measurements of physical objects. Their ability to generate and store secret information makes them a good candidate for security systems \cite{DDuanekey2019,assiri2019key}.
PUFs generate challenge and response pairs (CRPs); the challenge is created during the enrollment, whereas the response is generated each time authen- tications are needed. The authentication occurs when the number of mismatches between the challenge and response is low enough \cite{cambouPUFdesign,Assiri2021HomomorphicPM}. Memory structures SRAM, DRAM ReRAM, and MRAM are considered as good elements to generate PUFs \cite{cambouPUFdesign,gao2016emerging,Plusquellic2014}.

The keyless protocol that we studied used a memristor PUF, which has a lot  of active research, such as in the field of artificial intelligence (AI); the memristor is used to design artificial neurons or bioelectricity. The idea is that each cell should become a programmable node with its resistance adjusted to memorize learning patterns. Another field of active memristor research is to design resistive random-access memory (ReRAM), to replace mainstream mem ories, DRAM, and Flash \cite{johnsen2012introduction,tsai2008efficient,zhang2017artificial}. Moreover, it has been mentioned that the manufacturing of the ReRAMs creates natural variations amongst the cells; these variations can be exploited to design a PUF. Therefore, the challenge/response from  the ReRAM PUF is the resistance value obtained at a particular cell address,  where a specific current is injected \cite{cambou2020cryptography,li2017resistive,cambou2017ag,chen2015comprehensive}.

\subsection{Keyless Encryption}
As shown in \ref{fig:Thearchitectureschemesforkeylessencryption}, there are several steps that need to be taken to find the address to be read. The sender side will combine the random number and the password using the exclusive or (XOR). The outcome will be a binary stream of  64 bytes. Then the output of XOR will be used as the input of a hash function. The outcome of the hash function is 512 bits long according to \cite{cambou2020cryptography} and \cite{zhu2020extended}. But we can say that, according to the type of the hash function, we can know the size of the outcome. So, we are going to call the outcome of the hash function 512 bits long a short message digest (SMD). After getting the SMD, the sender side will try to extend it because the length of the message digest decides how many characters can be encrypted. To do this, the first n bits of SMD will be rotated;  each time, the output rotation will feed the Hash function to obtain a new SMD; and finally, all SMDs will be combined to get a longer message digest (LMD). For encryption, the message digest will be divided into n blocks. Each     block contains the address (7 bits) and current (3bits). The decimal value of the 7 bits will vary from 0 to 127, and the decimal value of 3 bits will vary from 0 to 7. The LMD will be generated from rotating the  first 16 bit to get 16 MDs, which is equal to 8196 bits. From there, the LMD are going to split to several blocks; if the size of each block equal 17 bits, then we can get approximately $8192 bits / 17 \approx 480$ blocks. Each block works with 2 characters, so  the length of the plaintext should be no more than 240 characters, unless we rotate more than 16 bits from the first MD. This protocol can be extended to a longer plaintext. Table \ref{tab:rotations} shows show how much n rotation can produce the
size of blocks and also the length of plaintext we can accept.in this study, we selected 16 rotations, which means the LMD will be 8192 bits; this  leads us to obtain 480 blocks.

The addressable table that we are referring to is a ReRAM PUF. The size of   the ReRAM data table is 128 by 8. The address defines which row, and the r         current defines which column to visit. The outcomes of the PUF will be used as  ciphertext to send over to the receiver. The encryption and decryption steps and how the keyless protocol works have been explained in more detail at \cite{cambou2020cryptography}.
However, in this study, we want to test the randomness of visiting the ReRAM PUF cell. It has been used for software protocol implementation of the TRNG  and hash function that have been recommended by NIST, but it is still not certain  if    the PUF cells are visited equally. \cite{cambou2020cryptography} mentions that if we encrypt with a new random number each time, we should visit different cells in the ReRAM PUF. So, this paper tries to infer the probability of visiting each cell in the PUF with a new random number each time. 

\begin{table}[]
\centering
\caption{Number of rotations needed to generate LMD based on the length of the plain text.}
\label{tab:rotations}
\begin{tabular}{|c|c|c|c|}
\hline
Maximum length of plain text & \# of blocks &\#  of rotation & Length of MD (bit) \\ \hline
240               & 480                & 16           & 8192                          \\ \hline
481               & 962               & 32           & 16384                          \\ \hline
963               & 1926               & 64          & 32768                          \\ \hline
\end{tabular}
\end{table}

\section{Objectives}
The issue with this protocol occurs when encrypting the exact plaintext multiple times. If we visit the same cells repeatedly, a hacker will use the frequency analysis technique to reveal the plaintext, and the security will be compromised. So, this study tries to determine if a frequency analysis attack can be exploited against a keyless encryption scheme using the ReRAM PUF protocol. In  the following, we will briefly describe the frequency analysis concept.

 \textbf{Frequency Analysis}
 Frequency analysis is studying the frequency of each letter in each language. For example, in English, “e” is the most frequently  used letter. Each language has its own proportions of appearance. Language characters are slightly different from each other. Therefore, texts written  in each language have certain common properties, which allow them to be distinguished from texts written in other languages. Each language has some  popular sequences of letters; for example, in English, there are often used vowels such as e, o, a, or the consonant t. Additionally, there are some very rare letters, for example, z or x. According to \cite{paar2009understanding}, the weakness of the block cipher is that the plaintext symbol usually assigns to the same ciphertext symbol, which means that the statistical features of plaintext are maintained in the ciphertext. Therefore, each time    we need to encrypt, the protocol generates a new RN that leads to a different MD from hash functions, so we can claim that with each new RN, the protocol  should visit different cells in the ReRAM PUF. In other words, the probability of visiting the cells in ReRAM PUF should be equal.

\section{Methods}
To accomplish this study, we ran a simulation using the encryption protocol, and we used statistical models to determine whether cells in the ReRAM PUF were disproportionately visited. This will help us understand whether hackers can exploit frequency analysis in this protocol. 
Knowing what the data looks like helps to figure out which is the best statistical model that should be chosen. This section will explain the method of data collection and data analysis. In addition, this section will show which statistical model was used.

\begin{table}[h]
\centering
\caption{Example of raw data for each run. Each cell has 8 different currents (I). The number shown in each cell is number of visits. }
\label{tab:RawData}
\resizebox{6cm}{!}{
    \begin{tabular}{c|c|c|c|c|c|c|c|c|}
    \cline{2-9}
    & $I_1$   & $I_2$   & $I_3$   & $I_4$  & $I_5$  & $I_6$  & $I_7$  & $I_8$  \\ \hline
    \multicolumn{1}{|c|}{$cell_1$}       & 0     & 0     & 1     & 3    & 0    & 4    & 1    & 0    \\ \hline
    \multicolumn{1}{|c|}{$cell_2$}       & 0     & 1     & 0     & 0    & 2    & 2    & 1    & 0    \\ \hline
    \multicolumn{1}{|c|}{$cell_3$}       & 0     & 2     & 1     & 0    & 0    & 0    & 3    & 0    \\ \hline
    \multicolumn{1}{|c|}{\vdots}       & \vdots     & \vdots    &\vdots     & \vdots    & \vdots    & \vdots    & \vdots    & \vdots    \\ \hline
    \multicolumn{1}{|c|}{$cell_{125}$}   & 1     & 0     & 2     & 4    & 0    & 0    & 0    & 1    \\ \hline
    \multicolumn{1}{|c|}{$cell_{126}$}   & 0     & 3     & 0     & 3    & 0    & 1    & 0    & 3    \\ \hline
    \multicolumn{1}{|c|}{$cell_{127}$}   & 1     & 0     & 0     & 4    & 1    & 0    & 0    & 4    \\ \hline
    \multicolumn{1}{|c|}{$cell_{128}$}   & 2     & 0     & 1     & 0    & 0    & 2    & 0    & 3    \\ \hline
    \end{tabular}
}
\end{table}

 \subsection{Data}
 We conducted an experiment in which we calculated the number of visits for each ReRAM cell with different current values. We used the same dataset that  has been used in \cite{cambou2020cryptography} and \cite{zhu2020extended}. The data is the output of reading the 128 cells with eight different currents. The PUF size that we have used is 128 cell $*$ 8  current $= 1024$  cells, as shown in table \ref{tab:RawData}. The value of each cell is the resistance value. We collected data each time we encrypted and decrypted because we counted each cell visited in ReRAM PUF.  Table \ref{tab:RawData} shows some cells were visited a few times, while others were not visited at all.  Table \ref{tab:RawData} shows an example for only visiting the ReRAM PUF cell for one run of encrypting and decrypting.
 
As mentioned earlier, this study would test the randomness of visiting each cell. According to \cite{cambou2020cryptography,zhu2020extended} the longest acceptable plaintext can be 240 characters.  Therefore, we loop the same plaintext several runs 10, 100, and 1000 runs. For each run, we count how many times each cell in ReRAM PUF is visited as  illustrated in table \ref{tab:datalabel} column 3. We frame the data as the number of times a specific cell is visited in a particular run of experiment, as illustrated in table \ref{tab:datalabel}.

\begin{table}[]
\centering
\caption{Data structure}
\label{tab:datalabel}
\resizebox{7cm}{!}{%
\begin{tabular}{|c|c|c|}
\hline
\textbf{Run Index} & \textbf{Cell Index} & \textbf{\begin{tabular}[c]{@{}c@{}}Number of times\\ the cell is visited\end{tabular}} \\ \hline
\rowcolor[HTML]{FFFFFF} 
{\color[HTML]{000000} \textbf{1}}                           & \textbf{1}    & \textbf{1}   \\ \hline
\rowcolor[HTML]{FFFFFF} 
{\color[HTML]{000000} \textbf{1}}                           & \textbf{2}    & \textbf{1}   \\ \hline
\rowcolor[HTML]{FFFFFF} 
{\color[HTML]{000000} \textbf{1}}                           & \textbf{3}    & \textbf{0}   \\ \hline
\rowcolor[HTML]{FFFFFF} 
{\color[HTML]{000000} \textbf{1}}                           & \textbf{4}    & \textbf{2}   \\ \hline
\rowcolor[HTML]{FFFFFF} 
\cellcolor[HTML]{EFEFEF}{\color[HTML]{000000} \textbf{...}} & \textbf{...}  & \textbf{...} \\ \hline
\rowcolor[HTML]{FFFFFF} 
\cellcolor[HTML]{C0C0C0}{\color[HTML]{000000} \textbf{2}}   & \textbf{1004} & \textbf{1}   \\ \hline
\rowcolor[HTML]{FFFFFF} 
\cellcolor[HTML]{C0C0C0}{\color[HTML]{000000} \textbf{2}}   & \textbf{1005} & \textbf{0}   \\ \hline
\rowcolor[HTML]{FFFFFF} 
\cellcolor[HTML]{9B9B9B}{\color[HTML]{000000} \textbf{...}} & \textbf{...}  & \textbf{...} \\ \hline
\rowcolor[HTML]{FFFFFF} 
\cellcolor[HTML]{656565}{\color[HTML]{000000} \textbf{10}}  & \textbf{991}  & \textbf{2}   \\ \hline
\rowcolor[HTML]{FFFFFF} 
\cellcolor[HTML]{656565}{\color[HTML]{000000} \textbf{10}}  & \textbf{992}  & \textbf{1}   \\ \hline
\rowcolor[HTML]{FFFFFF} 
\cellcolor[HTML]{656565}{\color[HTML]{000000} \textbf{10}}  & \textbf{993}  & \textbf{0}   \\ \hline
\end{tabular}%
}
\end{table}

\begin{figure}[t]
\vspace{2.5cm}
\centering
\includegraphics[width=0.9\linewidth]{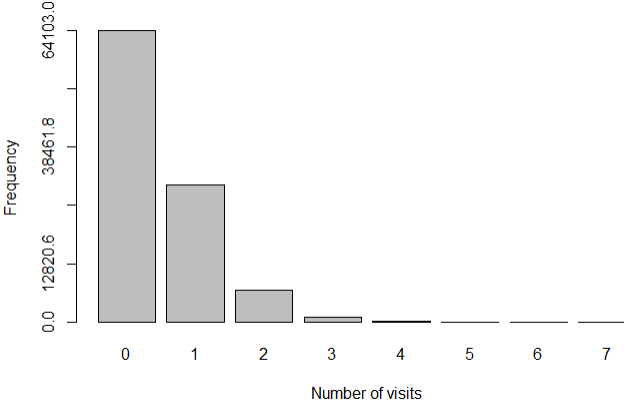}
\caption{The data distribution }
\label{fig:datadistribution}
\end{figure}

Each cell was visited a maximum of 7 times, but was much more likely to visited less than 3 times. See figure \ref{fig:datadistribution}.

\subsection{ Statistical Experiment}
To examine the randomness of the protocol, we conducted a regression to estimate the probability of visiting each cell in the ReRAM PUF. Specifically, we conducted a generalized linear mixed effects regression using a binomial error distribution. In this case, our goal was to determine the global average probability of visiting any specific cell, and to use random effects to determine if any particular cells had average visit probabilities that deviated significantly from the global average. 
In other words, if a cell had a random effect significantly different from zero, this would indicate a significantly higher or lower probability of visiting that cell compared to the average cell. 
Such a finding would provide evidence that the protocol is not sufficiently random to prevent frequency analysis. 
The model structure is as follows:

\begin{table}[]
\centering
\caption{Three model comparison. First model with two random effects. Second model with one random effect. Third model with no random effect. }
\label{tab:looic}
\resizebox{\textwidth}{!}{%
\begin{tabular}{c|c|c|c|}
\cline{2-4}
 & \begin{tabular}[c]{@{}l@{}}Estimated looic\\ with  standard error\end{tabular} & $elpd_{diff}$ & $se_{diff}$ \\ \hline
\multicolumn{1}{|l|}{\begin{tabular}[c]{@{}l@{}}Model with\\ two random effects\end{tabular}} & 18348 $\pm$ 152 & 0    & 0   \\ \hline
\multicolumn{1}{|l|}{\begin{tabular}[c]{@{}l@{}}Model with\\ one random effect\end{tabular}}  & 18346 $\pm$ 152 & 0.5 & 1.2 \\ \hline
\multicolumn{1}{|l|}{\begin{tabular}[c]{@{}l@{}}Model with\\ no random effect\end{tabular}}   & 18345 $\pm$ 153 & $-1.5$ & 1.2 \\ \hline
\end{tabular}%
}
\end{table}

\begin{eqnarray}
y_{c,r} \sim {Binomial}(n, \, P_{c,r})
\end{eqnarray}
\begin{itemize}
    \item $y_{c,r}$ represents the number of visits for each cell $c$ in a particular run of the experiment $r$.
    \item $n$ is the number of trials for each run, which is 480 for a 240 character message.
    \item $P_{c,r}$ is the probability of visiting each cell for each run.
\end{itemize}

We used a logit link with the following linear expression:

\begin{eqnarray}
logit(P_{c,r}) = \bar{a} + \eta_c + \eta_r
\end{eqnarray}

$\bar{a}$ is the global average probability of visiting each cell. $\eta_c$ is a random effect of cell $c$, the estimation of which helps us understand if we are more or less likely to visit a particular cell compared to the global average.  $\eta_r$ is a random effect of the experimental run. We included this random effect to account for any non-independence of the data from a particular run of the experiment. By definition, these random effects follow the assumption of being normally distributed:

\begin{eqnarray}
\eta_c \sim {Normal}(0, \, \sigma_c)
\end{eqnarray}
\begin{eqnarray}
\eta_r \sim {Normal}(0, \, \sigma_r)
\end{eqnarray}

We used two methods to evaluate if any cells deviated from the global average. First, we used model comparison to test whether the addition of the cell and run random effects were parsimonious. Specifically, we used the 
leave one out information criterion (LOO-IC) \cite{vehtari2017practical} to evaluate model performance between three competing models. 
The first model included both random effects, while second model included only the cell random effect, and the third model included no random effects (only the global average probability of visiting a cell) Table \ref{tab:looic}. 
If the model comparisons show that the simplest model (3) is the most parsimonious, then we have evidence that particular cells do not strongly deviate from the global average. 
However, because there are so many cells, this analysis may not be sufficiently robust to find a small number of cells with significant deviations.
Therefore, our second, complementary method was to evaluate whether the 95\% credible interval of each random cell effect overlapped zero. 
If the 95\% credible interval of a cell's random effect overlaps zero, then the probability of visiting that particular cell does not deviate significantly from the global average.
We also tested a more conservative 80\% credible interval.

We conducted this regression analysis using a Bayesian framework in the open-source statistical programming language, Stan \cite{carpenter2017stan}, via the R package \texttt{rstanarm} \cite{RCoreTeam1,Goodrichrstanarm,Stan2019}, and we used the \texttt{loo} \cite{Vehtari2019loo} package to estimate the LOO-IC of the competing models. 

\begin{figure}[]
\vspace{2.5cm}
\centering
\includegraphics[width=0.9\linewidth]{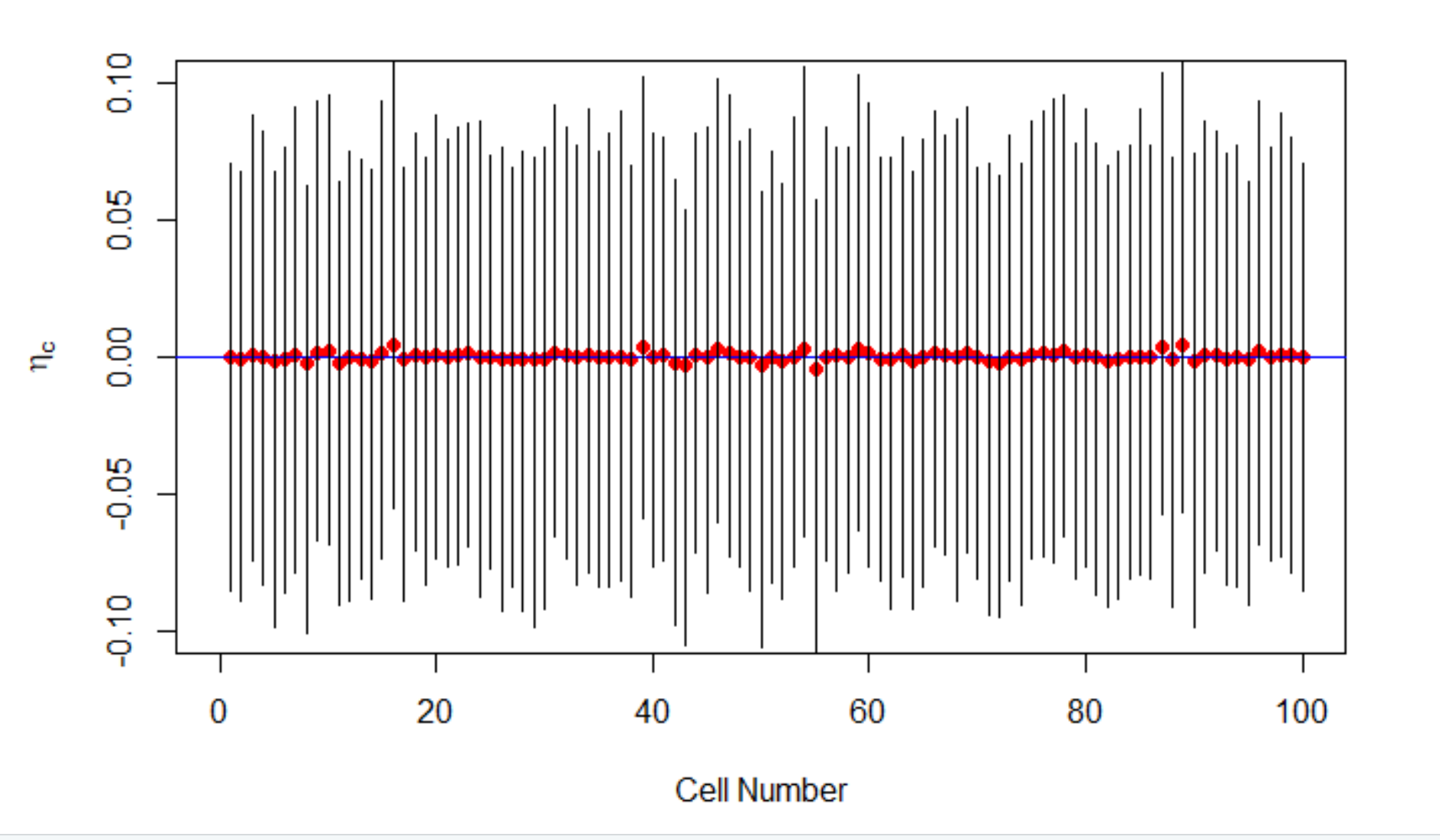}
\caption{Model with random cell effect.The error bar shows that none of the 95 percent credible intervals significantly deviate from 0. We can infer that there is no cell that has a big deviation. The highest value of cell deviation is almost 0.005 which is a insignificant value; In other words, it means that all probabilities for cells to be visited are close enough to the probability of the global mean.}
\label{fig:numeff}
\end{figure}

\section{Results}
Multiple lines of evidence suggest that cells do not significantly vary in the average probability of visit, providing evidence that the protocol is sufficiently random.
First, comparing the LOO-IC values of our statistical models showed that the random effects of cell and run did not add significant information to the model (Table \ref{tab:looic}). 
In other words, model comparison did not show that the probability of visiting specific cells deviated significantly from the global average.
More convincingly, none of the random cell effects had estimates that deviated significantly from zero based on the 95\% or 80\% credible intervals (see Figure \ref{fig:numeff} for a subset of these estimates).
This shows that the model was not able to detect a single cell with significant deviation.
The global probability of visiting any specific cell was of course very low. The global intercept $\bar{a}$ median and 95\% credible interval on the logit scale was -6.9 which is equal to 0.001006771 on the probability scale.

The results of this study therefore provide evidence that visiting cells is sufficiently random. 
It is possible that if we ran the experiment millions of times, that we could detect much smaller deviations in the probability of visiting each cell, but we do not suspect that such small deviations would benefit a hacker. 
We therefore claim that a hacker will not be able to use a frequency analysis attack to break the ciphertext. 
In addition, our study further suggests that it is secure to use the ReRAM PUF for keyless encryption. This might be one big step to enhance the security of IoT.

\section{Conclusion}

Keyless protocol based on ReRAM PUF shows it is sufficiently secure during encryption and decryption. Using PUFs has obstacles that need to be considered. One of these obstacles is visiting the cells; studying the randomness  of visiting the cells and exploring the probability of visiting the cells indicates the ReRAM PUF helps eliminate the risk of using frequency analysis attacks against the protocol. In general, visiting the same cell several times during the encryption phase makes the frequency analysis attack work against the keyless encryption protocol based on ReRAM PUF. In this study, we ran a statistical model to test the probability of visiting the cells; the binomial model is applied in this study since we aim to identify the success or failure of each cell. The outcome of the statistical experiment illustrates that visiting cells is entirely random. Therefore, using a frequency analysis attack is not effective against keyless encryption protocol based on a ReRAM PUF.

\textbf{Acknowledgments}.
The authors are thanking the contribution of several graduate students at the
cyber-security lab at Northern Arizona University, in particular, Christopher Philabaum, Ian Burke, and Alireza Shamsoshoara. Also, the author is thanking the contribution
of Jazan University.

\bibliography{ref}

\begin{thebibliography}{10}

\bibitem{cambou2021post}
B.~Cambou, M.~Gowanlock, B.~Yildiz, D.~Ghanaimiandoab, K.~Lee, S.~Nelson,
  C.~Philabaum, A.~Stenberg, and J.~Wright, ``Post quantum cryptographic keys
  generated with physical unclonable functions,'' {\em Applied Sciences},
  vol.~11, no.~6, p.~2801, 2021.

\bibitem{keshavarz2018towards}
M.~Keshavarz and M.~Anwar, ``Towards improving privacy control for smart homes:
  A privacy decision framework,'' in {\em 2018 16th Annual Conference on
  Privacy, Security and Trust (PST)}, pp.~1--3, IEEE, 2018.

\bibitem{cambou2020cryptography}
B.~Cambou, D.~H{\'e}ly, and S.~Assiri, ``Cryptography with analog scheme using
  memristors,'' {\em ACM Journal on Emerging Technologies in Computing Systems
  (JETC)}, vol.~16, no.~4, pp.~1--30, 2020.

\bibitem{roman2011securing}
R.~Roman, P.~Najera, and J.~Lopez, ``Securing the internet of things,'' {\em
  Computer}, vol.~44, no.~9, pp.~51--58, 2011.

\bibitem{baracaldo2016securing}
N.~Baracaldo, L.~A.~D. Bathen, R.~O. Ozugha, R.~Engel, S.~Tata, and H.~Ludwig,
  ``Securing data provenance in internet of things (iot) systems,'' in {\em
  International conference on service-oriented computing}, pp.~92--98,
  Springer, 2016.

\bibitem{kumar2003keyless}
S.~Kumar and V.~Kumar, ``Keyless encryption of messages using challenge
  response,'' Mar.~18 2003.
\newblock US Patent 6,535,980.

\bibitem{chandrasekaran2015keyless}
S.~Chandrasekaran, S.~Murugan, A.~C. Ramachandran, and L.~Velusamy, ``Keyless
  challenge and response system,'' Feb.~24 2015.
\newblock US Patent 8,966,254.

\bibitem{chua1971memristor}
L.~Chua, ``Memristor-the missing circuit element,'' {\em IEEE Transactions on
  circuit theory}, vol.~18, no.~5, pp.~507--519, 1971.

\bibitem{johnsen2012introduction}
G.~K. Johnsen, ``An introduction to the memristor-a valuable circuit element in
  bioelectricity and bioimpedance,'' {\em Journal of Electrical Bioimpedance},
  vol.~3, no.~1, pp.~20--28, 2012.

\bibitem{stanley2013we}
R.~Stanley~Williams, ``How we found the missing memristor,'' in {\em Chaos,
  CNN, Memristors and Beyond: A Festschrift for Leon Chua With DVD-ROM,
  composed by Eleonora Bilotta}, pp.~483--489, World Scientific, 2013.

\bibitem{adam20173d}
G.~C. Adam, B.~Chrakrabarti, H.~Nili, B.~Hoskins, M.~A. Lastras-Monta{\~n}o,
  A.~Madhavan, M.~Payvand, A.~Ghofrani, K.-T. Cheng, L.~Theogarajan, {\em
  et~al.}, ``3d reram arrays and crossbars: Fabrication, characterization and
  applications,'' in {\em 2017 Ieee 17th International Conference on
  Nanotechnology (Ieee-Nano)}, pp.~844--849, IEEE, 2017.

\bibitem{cambou2020puf}
B.~F. Cambou, R.~C. Quispe, and B.~Babib, ``Puf with dissolvable conductive
  paths,'' May~28 2020.
\newblock US Patent App. 16/493,263.

\bibitem{zhu2020extended}
Y.~Zhu, B.~Cambou, D.~Hely, and S.~Assiri, ``Extended protocol using keyless
  encryption based on memristors,'' in {\em Science and Information
  Conference}, pp.~494--510, Springer, 2020.

\bibitem{paar2009understanding}
C.~Paar and J.~Pelzl, {\em Understanding cryptography: a textbook for students
  and practitioners}.
\newblock Springer Science \& Business Media, 2009.

\bibitem{LSORainb88:online}
``Lso-rainbowcrack.doc.''
  \url{http://www.windowsecurity.com/uplarticle/Cryptography/LSO-RainbowCrack.pdf}.
\newblock (Accessed on 06/30/2021).

\bibitem{DDuanekey2019}
D.~D. Booher, B.~Cambou, A.~H. Carlson, and C.~Philabaum, ``Dynamic key
  generation for polymorphic encryption,'' in {\em 2019 IEEE 9th Annual
  Computing and Communication Workshop and Conference (CCWC)}, pp.~0482--0487,
  2019.

\bibitem{assiri2019key}
S.~Assiri, B.~Cambou, D.~D. Booher, D.~G. Miandoab, and M.~Mohammadinodoushan,
  ``Key exchange using ternary system to enhance security,'' in {\em 2019 IEEE
  9th Annual Computing and Communication Workshop and Conference (CCWC)},
  pp.~0488--0492, IEEE, 2019.

\bibitem{cambouPUFdesign}
B.~Cambou and M.~Orlowski, ``Puf designed with resistive ram and ternary
  states,'' in {\em Proceedings of the 11th Annual Cyber and Information
  Security Research Conference}, CISRC '16, (New York, NY, USA), Association
  for Computing Machinery, 2016.

\bibitem{Assiri2021HomomorphicPM}
S.~Assiri and B.~Cambou, ``Homomorphic password manager using multiple-hash
  with puf,'' 2021.

\bibitem{gao2016emerging}
Y.~Gao, D.~C. Ranasinghe, S.~F. Al-Sarawi, O.~Kavehei, and D.~Abbott,
  ``Emerging physical unclonable functions with nanotechnology,'' {\em IEEE
  access}, vol.~4, pp.~61--80, 2016.

\bibitem{tsai2008efficient}
J.-L. Tsai, ``Efficient multi-server authentication scheme based on one-way
  hash function without verification table,'' {\em Computers \& Security},
  vol.~27, no.~3-4, pp.~115--121, 2008.

\bibitem{zhang2017artificial}
X.~Zhang, W.~Wang, Q.~Liu, X.~Zhao, J.~Wei, R.~Cao, Z.~Yao, X.~Zhu, F.~Zhang,
  H.~Lv, {\em et~al.}, ``An artificial neuron based on a threshold switching
  memristor,'' {\em IEEE Electron Device Letters}, vol.~39, no.~2,
  pp.~308--311, 2017.

\bibitem{li2017resistive}
H.~Li, T.~F. Wu, S.~Mitra, and H.-S.~P. Wong, ``Resistive ram-centric
  computing: Design and modeling methodology,'' {\em IEEE Transactions on
  Circuits and Systems I: Regular Papers}, vol.~64, no.~9, pp.~2263--2273,
  2017.

\bibitem{cambou2017ag}
B.~Cambou, F.~Afghah, D.~Sonderegger, J.~Taggart, H.~Barnaby, and M.~Kozicki,
  ``Ag conductive bridge rams for physical unclonable functions,'' in {\em 2017
  IEEE International Symposium on Hardware Oriented Security and Trust (HOST)},
  pp.~151--151, IEEE Computer Society, 2017.

\bibitem{chen2015comprehensive}
A.~Chen, ``Comprehensive assessment of rram-based puf for hardware security
  applications,'' in {\em 2015 IEEE International Electron Devices Meeting
  (IEDM)}, pp.~10--7, IEEE, 2015.

\bibitem{vehtari2017practical}
A.~Vehtari, A.~Gelman, and J.~Gabry, ``Practical bayesian model evaluation
  using leave-one-out cross-validation and waic,'' {\em Statistics and
  computing}, vol.~27, no.~5, pp.~1413--1432, 2017.

\bibitem{carpenter2017stan}
B.~Carpenter, A.~Gelman, M.~D. Hoffman, D.~Lee, B.~Goodrich, M.~Betancourt,
  M.~Brubaker, J.~Guo, P.~Li, and A.~Riddell, ``Stan: A probabilistic
  programming language,'' {\em Journal of statistical software}, vol.~76,
  no.~1, pp.~1--32, 2017.

\bibitem{RCoreTeam1}
{R Core Team}, {\em R: A Language and Environment for Statistical Computing}.
\newblock R Foundation for Statistical Computing, Vienna, Austria, 2019.

\bibitem{Goodrichrstanarm}
B.~Goodrich, J.~Gabry, I.~Ali, and S.~Brilleman, ``rstanarm: {Bayesian} applied
  regression modeling via {Stan}.,'' 2018.
\newblock R package version 2.17.4.

\bibitem{Stan2019}
{Stan Development Team}, ``{RStan}: the {R} interface to {Stan},'' 2019.
\newblock R package version 2.19.2.

\bibitem{Vehtari2019loo}
A.~Vehtari, J.~Gabry, Y.~Yao, and A.~Gelman, ``loo: Efficient leave-one-out
  cross-validation and waic for bayesian models,'' 2019.
\newblock R package version 2.1.0.

\end{thebibliography}
\bibliographystyle{ieeetr}

\end{document}